\documentstyle[epsf,twocolumn,prl,aps]{revtex}
\begin{document}
%
\draft
%
\wideabs{
\title{Quasi-free Compton Scattering from the Deuteron and Nucleon 
Polarizabilities}
\author{N.R.~Kolb, A.W.~Rauf, R.~Igarashi, D.L.~Hornidge, R.E.~Pywell,
B.J. Warkentin}
\address{Saskatchewan Accelerator Laboratory,
University of Saskatchewan,
Saskatoon, SK, Canada, S7N 5C6}
\author{E. Korkmaz}
\address{Department of Physics, University of Northern British Columbia,
Prince George, BC, Canada, V2N 4Z9}
\author{G.~Feldman, G.V.~O'Rielly}
\address{Department of Physics, The George Washington University, Washington, 
D.C. 20052}
\date{\today}
\maketitle
%
\begin{abstract}
Cross sections for quasi-free Compton scattering from the deuteron were 
measured for incident energies of E$_\gamma$=236--260~MeV
at the laboratory angle $\theta_{\gamma^\prime}= {-135^\circ}$.  The 
recoil nucleons were detected in a liquid-scintillator array situated at 
$\theta_N = {20^\circ}$.
The measured differential cross sections were used, with 
the calculations of Levchuk {\it et al.}, to determine
the polarizabilities of the bound nucleons.
For the bound proton, the extracted values were consistent
with the accepted value for the free proton.  
Combining our results for the bound neutron with those
from Rose {\it et al.}, we obtain one-sigma constraints of
$\overline{\alpha}_n = 7.6-14.0$ and 
$\overline{\beta}_n = 1.2-7.6$.

\end{abstract}

\pacs{PACS numbers: 25.20.Dc, 13.40.Em, 13.60.Fz, 14.20.Dh}

}

The electric $\overline{\alpha}$ and magnetic $\overline{\beta}$
polarizabilities constitute the first-order responses of the internal
structure of a nucleon to externally applied electric and magnetic
fields.  Compton scattering from the proton has been used
extensively to determine the polarizabilities of the proton
(see Ref.~\cite{Mac-95} and references contained therein).  Such
measurements are sensitive to the sum (difference) of the polarizabilities
for photons scattered at forward (backward) angles.  
Ref.~\cite{Mac-95} reports the experimental status of 
the proton polarizabilities, in the usual
units of $10^{-4}$~fm$^3$:
\begin{eqnarray}
	(\overline{\alpha}-\overline{\beta})_p & = & 10.0\pm1.5\pm0.9,
	\label{prot-diff}\\
	(\overline{\alpha}+\overline{\beta})_p & = & 15.2\pm2.6\pm0.2,
\end{eqnarray}
where the first error is the combined statistical and systematic, and the
second is due to the model dependence of the dispersion-relation
extraction method.

Disperson sum rules relate the sum of the polarizabilities to the nucleon
photoabsorption cross section.  The generally accepted results
\cite{Pet-81} were
\begin{eqnarray}
	(\overline{\alpha}+\overline{\beta})_p & = & 14.2 \pm 0.5, 
	\label{prot-sum} \\
	(\overline{\alpha}+\overline{\beta})_n & = & 15.8 \pm 0.5.
\end{eqnarray}
There have been recent reevaluations of these sum rules which yield
$13.69 \pm 0.14$ ($14.40 \pm 0.66$) \cite{Bab-98}
and $14.0\pm 0.5$ ($15.2\pm 0.5$) \cite{levchuk1} 
for the proton (neutron).  
The proton polarizabilities, obtained from Eqs.~\ref{prot-diff} and
\ref{prot-sum}, are
\begin{equation}
	\overline{\alpha}_p = 12.1\pm 0.8\pm 0.5,   
	\hspace{0.5cm}
	\overline{\beta}_p  =  2.1\mp 0.8\mp 0.5.
	\label{prot-polar}
\end{equation}

The status of the neutron polarizabilities is much less satisfactory
(see Ref.~\cite{Wiss-98}  and references contained therein).
Measurements of the electric polarizability
of the neutron have been done by low-energy neutron scattering
from the Coulomb field of a heavy nucleus.  The extracted
values fall in the range $\alpha_n$=0--19 \cite{Sch-91,Koe-95,Enik-97}.

Elastic Compton scattering from the deuteron has also been used to 
extract information
on the nucleon polarizabilities.  Using their theoretical model, 
Levchuk and L'vov \cite{levchuk1} have reported values of
\begin{equation}
(\overline{\alpha}-\overline{\beta})_n = -2 \pm 3, 
	\hspace{0.5cm}
(\overline{\alpha}+\overline{\beta})_n = 20 \pm 3, \label{neut-diff} 
\end{equation}
from fitting to the data of Refs.~\cite{Lucas-94,Horn-99}. 
The large discrepancy between $(\overline{\alpha}-\overline{\beta})_n$
and $(\overline{\alpha}-\overline{\beta})_p$
(Eq.~\ref{prot-diff}) may be indicative of shortcomings 
in the theoretical models used to extract the polarizabilities
from the $d(\gamma,\gamma)d$ reaction.

The quasi-free Compton scattering reaction 
$d(\gamma,\gamma^\prime n)p$, in which the scattered photon is
detected in coincidence with the recoil neutron, can be used
to minimize the model dependence of the extracted polarizabilities. 
There has been one measurement reported on this reaction using
bremsstrahlung photons with an endpoint of 130 MeV~\cite{Rose-90}.  
A most-probable value of $\overline{\alpha}_n=10.7$ was obtained
with an upper limit of 14.0 but with no constraint on the lower limit.
In addition, the model dependence is strong at these lower energies.  

Levchuk {\it et al.} \cite{Lev-94} have determined that, within the 
context of their
model, the sensitivity to the neutron amplitude is maximized (and
model dependence minimized) for E$_\gamma$=200--300~MeV
and backward angles for the scattered photons. 
For $E_\gamma$=247~MeV and 
$\theta_{\gamma^\prime}$=$-135^\circ$,
the contribution of the spectator nucleon modifies the cross section
by only $\sim$4\% at the quasi-free peak.
Furthermore, the free-neutron cross section can be related to the 
cross section in the center of the neutron quasi-free peak via a 
spectator formula \cite{Lev-94,Wiss-99}
\begin{equation}
	\frac{ d\sigma(\gamma n\rightarrow \gamma^\prime n) }
	     { d\Omega_{\gamma^\prime} } = 
	\frac{ (2\pi)^3 }{ u^2(0) }
	\frac{ E_\gamma E_{\gamma^\prime} }
	     { |{\bf p}_n|m{E_{\gamma^\prime}^{(n)2}} }
	\frac{ d^3\sigma(\gamma d\rightarrow \gamma^\prime np) }
	     {d\Omega_{\gamma^\prime}d\Omega_n dE_n },
	\label{spectator}
\end{equation}
where $u(0)$ is the S-wave amplitude of the deuteron wave function
at zero momentum, and $E_{\gamma^\prime}^{(n)}$
is the photon energy in the rest frame of the final neutron.
The cross section on the right side of the
equation must be corrected for the small contribution of the
spectator proton (its pole diagram and final state 
interactions).

The present measurement was performed at the Sask\-atch\-ewan
Accelerator Laboratory (SAL).  
The $d(\gamma,\gamma^\prime n)p$ and $d(\gamma,\gamma^\prime p)n$
cross sections were measured simultaneously, in kinematics that 
emphasized the quasi-free reaction.   Data were also obtained
from the free proton, using the same apparatus, for calibration
and normalization purposes.
An electron beam of 292~MeV and {$\sim$60\%} duty factor 
produced bremsstrahlung photons, which were tagged 
in the energy range 236--260~MeV, with a resolution of 0.4~MeV.
The average, integrated tagged flux was $\sim{6}\times{10^{6}}$ 
photons/s with a tagging efficiency of $\sim$45\%.
The total integrated luminosity for the deuterium
measurements was $1.25\times 10^{37}$~photons/cm$^{2}$.

The cryogenic target cell was a vertical cylinder (12~cm in diameter), 
with Mylar walls, containing liquid hydrogen ($p(\gamma,\gamma)$)
or liquid deuterium ($d(\gamma,\gamma^\prime N$) and
$d(\gamma,pn)$). 
The quasi-free scattering of 247~MeV photons to 
$-135^\circ$ from deuterium results in
the recoil nucleon being confined to a forward cone
centered around 18$^\circ$.  These nucleons 
have a central kinetic energy of $\sim$77~MeV and were 
detected in
a liquid-scintillator array \cite{Kork-99} situated
at 20$^\circ$, subtending a solid angle of approximately 0.11~sr.
The array consisted of 85 Lucite-walled cells, filled with
BC-505 liquid scintillator.  Between the cells and the target
were thin plastic scintillators which acted as veto detectors
for neutrons or as $\Delta$E detectors for protons.

The neutron efficiency of the array was measured at the beginning
of the experiment via the $d(\gamma,pn)$ reaction.  A 
plastic-scintillator $\Delta$E$\cdot$E telescope was placed at
80$^\circ$ to detect the protons while the neutron
array was located at $-77^\circ$.  The detection efficiency
was measured for neutrons with kinetic energies of 50--100~MeV,
covering the range of interest for the quasi-free reaction.
With the applied threshold of 11~MeV (6~MeV{\it electron 
equivalent}) the efficiency ranged from 5.2--5.9\%.

Photons scattered from the target were 
detected in the large-volume Boston University NaI (BUNI) gamma-ray 
spectrometer \cite{Miller-88}.
BUNI is composed of five optically-isolated segments of NaI,
the core and four surrounding quadrants.
Plastic-scintillator detectors were used for rejection of cosmic
rays as well as charged particles from the target.
BUNI was set at $-135^\circ$ and subtended a solid angle of 
0.013~sr.  Zero-degree calibrations of
BUNI were done at the beginning and end of the experiment,
in order to obtain both the lineshape of BUNI and an energy
calibration for the NaI core.  The NaI quadrants were
calibrated periodically with a radioactive source (Th-C).

\begin{figure}[htb]
\begin{center}
\epsfxsize=3.38in
\epsffile{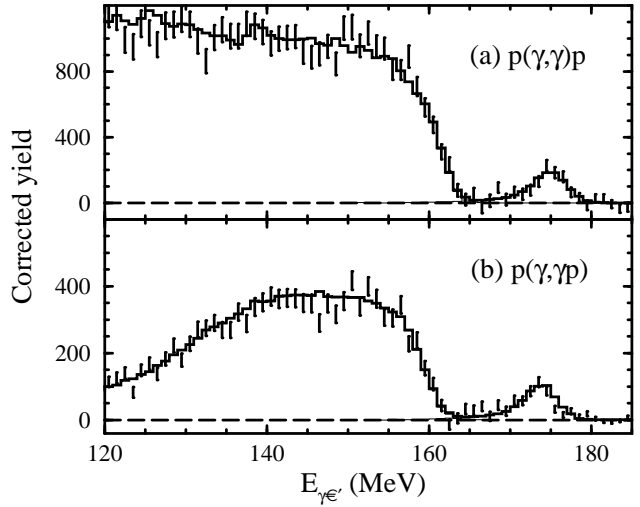}
\caption{Detected photon energy spectra 
for E$_\gamma = 247$~MeV and $\theta_\gamma=-135^\circ$
from the free proton without (a)
and with (b) detection of the proton.  The histograms are from a
simulation of the experiment.}
\label{proton}
\end{center}
\end{figure}

Figure~\ref{proton} depicts the BUNI energy spectrum 
(E$_{\gamma^\prime}$), corrected for photon absorption 
effects ($\sim$3\%), from the free proton.  
Random coincidences as well as contributions from empty-target
backgrounds (36\% for $(\gamma,\gamma)$ and 5\% for $(\gamma,\gamma p)$)
have been subtracted.
The spectra are summed over all tagged photon energies and shifted 
to the maximum incident energy of 260 MeV, with the appropriate
kinematic corrections to account for the different scattered photon
energy at each incident photon energy.

Figure~\ref{proton}(a) is the energy spectrum from a free proton,
without detection of the recoil proton.  The Compton scattered events
(the peak near 175~MeV) are clearly distinguishable from the large
number of events from neutral pion ($\pi^0$) production 
(below $\sim$165~MeV).
The histogram is the result of a simulation 
of the experiment.
The final yield was determined by integrating the data
from 170 to 190~MeV and correcting for the
Compton events excluded from this region (18\%) and the
$\pi^0$ events included in this region (1\%).  
For E$_\gamma = 247$~MeV and $\theta_\gamma=-135^\circ$,
the cross section for elastic Compton scattering 
from the free proton ($p(\gamma,\gamma)p$) was determined to be 
\begin{equation}
	\frac{d\sigma}{d\Omega_\gamma} = 94.6 \pm 9.0 \pm 3.8~{\rm nb/sr}, 
	\label{free}
\end{equation}
where the first error is statistical and the second is
systematic.  This is in good agreement with recent 
measurements \cite{Wiss-99,Hallin,legs}.
The dispersion calculations
of L'vov {\it et al.} \cite{Lvov-97} predict a somewhat lower value 
of 78.0~nb/sr using
$(\overline{\alpha}-\overline{\beta})_p$=10.

The efficiency for the detection of protons in the array
was determined in situ
from the fraction of the $p(\gamma,\gamma)p$ events
(Fig.~\ref{proton}(a)) that
included a detected proton (Fig.~\ref{proton}(b)).
This correction ($0.58 \pm 0.02$) was accounted for by a combination
of the simulation and cuts in the analysis.
\begin{figure}[htb]
\begin{center}
\epsfxsize=3.38in
\epsffile{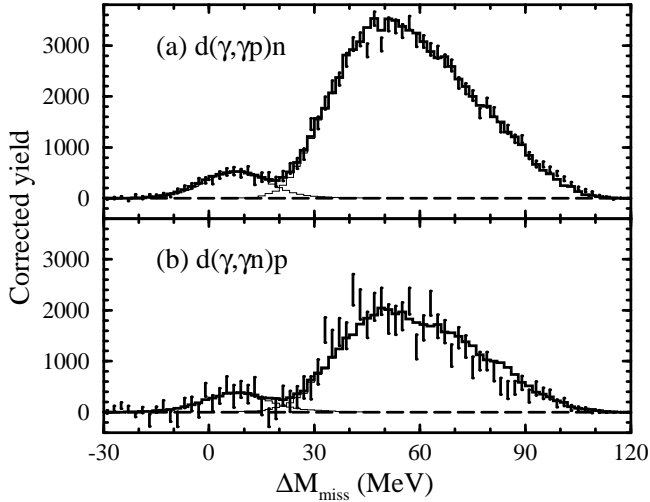}
\caption{Missing mass ($\Delta {\rm M_{miss} = M_X-M_N}$) spectra
for E$_\gamma = 247$~MeV and $\theta_{\gamma^\prime}=-135^\circ$
from the bound proton (a)
and the bound neutron (b). The histograms are from simulations of the
experiment.}
\label{miss-mass}
\end{center}
\end{figure}

For Compton scattering from a bound nucleon,
the Fermi motion causes overlap in E$_{\gamma^\prime}$
between the Compton and $\pi^0$ events.  
In order to obtain better separation,
the missing mass (${\rm M_X}$) of the spectator nucleon
was calculated.  Figure~\ref{miss-mass} depicts the difference
between the missing mass and the spectator mass 
($\Delta {\rm M_{miss}}$)
for the two quasi-free reactions.  
The yields are corrected for random coincidences, empty-target
backgrounds (12--14\%), and detection efficiencies.
The histograms are from simulations of the experiment.  
The thin lines are the individual
contributions from the $\pi^0$ and Compton reactions
while the thick line is the sum.
The simulation assumes that
the deuteron reactions proceed via quasi-free kinematics.  The spectator
nucleon is distributed according to its Fermi
distribution and the angular distribution of the $\pi^0$ or photon is 
determined by the theoretical calculations of Levchuk {\it et al.}
The simulation gives a good fit to the data, and is
relatively insensitive to the input angular distribution ($\sim$2\%).
The $\chi^2/N_{d.o.f.}$ is 1.4 (proton) and 1.5 (neutron) for 
$N_{d.o.f.}$=63 ($-20$ to 110~MeV).
To extract the yields for the Compton
reactions, the data were integrated from $-20$ to 
+20~MeV and corrected for the
Compton events excluded from this region (10--15\%) and the
$\pi^0$ events included in this region (7--8\%).  

In order to compare to theoretical predictions, 
the measured cross sections were interpolated to the quasi-free 
kinematic point ($P_{spectator}=0$) through use of the simulation.  
The average and root-mean-square values for $P_{spectator}$
were 48 and 55~MeV/c, respectively.
For E$_\gamma = 247$~MeV and $\theta_{\gamma^\prime}=-135^\circ$,
the final differential cross 
sections, at the quasi-free point, are
\begin{eqnarray}
\frac{d^3\sigma}{d\Omega_\gamma d\Omega_N dE_N} & = & 
    		56.5 \pm 2.8 \pm 4.5~({\rm proton}),\label{qf-prot}\\
	& = & 	33.3 \pm 7.2 \pm 3.0~({\rm neutron}),\label{qf-neut}
\end{eqnarray}
where the units are nb/sr$^2$/MeV, the first error is statistical, 
and the second is systematic.
The proton result is consistent with the recent results of
Wissmann {\it et al.} \cite{Wiss-99}.  

\begin{figure}[htb]
\begin{center}
\epsfxsize=3.38in
\epsffile{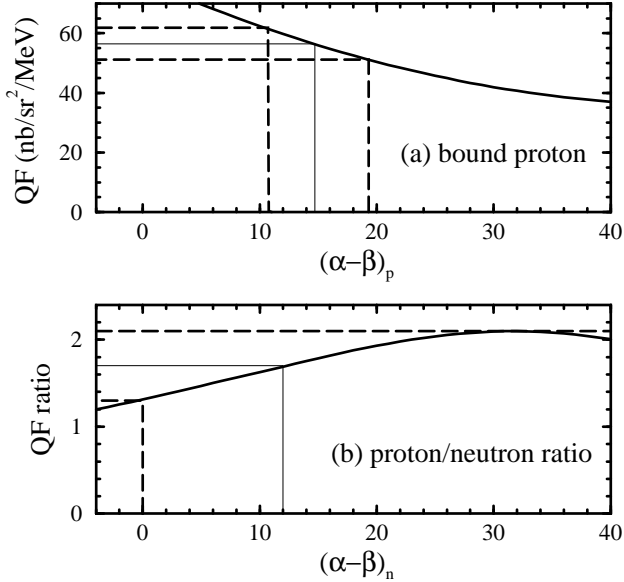}
\caption{(a) Quasi-free cross section for the bound proton.
(b) Ratio of the quasi-free cross section of the bound
proton and neutron.  The curves are the theoretical calculation 
of Levchuk {\it et al.}}
\label{ratio}
\end{center}
\end{figure}

Since the back-angle Compton cross section is primarily sensitive to 
the difference of the polarizabilities,  
$(\overline{\alpha}-\overline{\beta})_p$ was used as a fit parameter in
matching the theoretical calculation Levchuk {\it et al.} \cite{Lev-94}
to the bound proton data (Fig.~\ref{ratio}(a)).
The solid line is the theoretical 
calculation, the thin line is the central fitted 
value, and the horizontal dashed lines
indicate the combined-error band (the quadratic sum of statistical and
systematic).  The model dependence of the calculation is small compared
to the experimental errors.  The fitted value,
\begin{equation}
	(\overline{\alpha}-\overline{\beta})_p = {14.7^{+4.6}_{-4.0}},
\end{equation}
is consistent with the free proton result (Eq.~\ref{prot-diff}),
as well as with the quasi-free results of Wissmann {\it et al.} 
\cite{Wiss-99}. 
This indicates that the theoretical calculations are 
adequately describing the nuclear effects.

The theoretical calculations have been carried out using a
`traditional' value for the backward-spin polarizability of
$\gamma_\pi$=$-36.8$.  Using the value $\gamma_\pi$=$-27.1$, suggested
by Tonnison {\it et al.} \cite{tonnison}, results in a fitted value of
\begin{equation}
	(\overline{\alpha}-\overline{\beta})_p = 26.6^{+7.8}_{-5.8},
\end{equation}
which is over two standard deviations from the free value.  Our
data show no evidence for modification of $\gamma_\pi$, in
agreement with the conclusions of Wissmann {\it et al.}
\cite{Wiss-99}. 

The ratio of the bound proton to the bound neutron 
quasi-free cross section is 1.70$\pm$0.38$\pm$0.13.
Using the ratio of the measured cross sections
minimizes the effect of systematic errors on 
the extraction of $(\overline{\alpha}-\overline{\beta})_n$,
and reduces the model dependence inherent in
extracting information on the free-neutron from quasi-free
results.  The ratio for the theoretical
calculation was obtained by fixing the proton polarizabilities
(Eq.~\ref{prot-diff}) and treating 
$(\overline{\alpha}-\overline{\beta})_n$ as a fit parameter
(Fig.~\ref{ratio}(b)).  
The best fit was obtained with
$(\overline{\alpha}-\overline{\beta})_n = 12$, with a lower
limit of $0$.  No upper limit can be obtained.
The model uncertainty in extracting 
$(\overline{\alpha}-\overline{\beta})_n$
is expected to be negligible compared to the experimental error.
The central value is markedly different from the value extracted 
from elastic Compton scattering from the deuteron (Eq.~\ref{neut-diff}).
However, the two are within errors.
Invoking the sum rule \cite{levchuk1}, the values for the polarizabilities
are $\overline{\alpha}_n = 13.6$, with a lower limit of 7.6,
and $\overline{\beta}_n$ = 1.6, with an upper limit of 7.6.
The upper (lower) limits on 
$\overline{\alpha}_n$ ($\overline{\beta}_n$)
are unconstrained
by the present measurement.
Combining our results with those
from Rose {\it et al.} \cite{Rose-90}, we obtain one-sigma constraints of
$\overline{\alpha}_n = 7.6-14.0$ and 
$\overline{\beta}_n = 1.2-7.6$.
These can be compared with recent Chiral Perturbation Theory (ChPT)
predictions at $O(Q^4)$ \cite{bernard}
\begin{equation}
	\overline{\alpha}_n  =  13.4 \pm 1.5, \hspace{1cm}
	\overline{\beta}_n   =  7.8 \pm 3.6. 
	\label{neut-polar}
\end{equation}

From Eqs.~\ref{spectator} and \ref{qf-prot} the effective 
free-proton cross section at $\theta_{\gamma^\prime}=-135^\circ$
can be determined from that of the bound proton
($d(\gamma,\gamma^\prime p)n$).
The result is 
\begin{equation}
	\frac{ d\sigma(\gamma p\rightarrow \gamma^\prime p) }
	     { d\Omega_\gamma } = 66.8 \pm 3.4 \pm 5.3~{\rm nb/sr},
\end{equation}
corresponding to E$_\gamma=244.6$~MeV for the free
proton.  Adjusting Eq.~\ref{free} to this
lower energy results in a $\sim$5\% drop in the measured cross section
to 90.1$\pm$8.6$\pm$3.6~nb/sr. 
The ratio of the measured to effective free values is
$1.35\pm 0.18$, indicating they are not in good agreement.  
Model-dependence in extracting effective free-neutron values 
from quasi-free should be minimized by using the ratio of quasi-free 
cross sections, as discussed above.
The dispersion calculations of L'vov {\it et al.} \cite{Lvov-97} predict 
a free value of 74.2~nb/sr for $(\overline{\alpha}-\overline{\beta})_p$=10.

Eqs.~\ref{spectator} and \ref{qf-neut} can also be used to 
determine the effective free cross section for the neutron at
$\theta_{\gamma^\prime}=-135^\circ$:
\begin{equation}
	\frac{ d\sigma(\gamma n\rightarrow \gamma^\prime n) }
	     { d\Omega_\gamma } = 39.5 \pm 8.5 \pm 3.6~{\rm nb/sr},
\end{equation}
corresponding to E$_\gamma=244.6$~MeV for the free neutron.
This is the first
experimental determination of the cross section for Compton scattering
from the free neutron.
An alternate method for extracting the free-neutron cross section from
the measured values is to take the ratio of free to quasi-free cross
sections for the proton (Eqs.~\ref{free} and \ref{qf-prot})
and multiply by the quasi-free neutron cross section (Eq.~\ref{qf-neut}),
resulting in 55.8$\pm$13.5$\pm$4.9.  
The ratio of the extracted free values $1.41\pm 0.47$
is consistent with unity, albeit with large errors.
The dispersion calculations of L'vov {\it et al.} \cite{Lvov-97} predict 
a free value of 45.7~nb/sr (for $(\overline{\alpha}-\overline{\beta})_n$=10)
in agreement with the experimental results.

In summary,
the polarizabilities of the bound proton and neutron have,
for the first time, been simultaneously extracted from
quasi-free Compton scattering measurements.  The values
for the neutron (Eq.~\ref{neut-polar}) are consistent with those known
for the free proton (Eq.~\ref{prot-polar}), as expected from 
charge-symmetry arguments, and are in accord with recent ChPT
predictions.

The authors would like to thank M.I.~Levchuk, A.I. L'vov, and 
V.A. Petrun'kin for supplying the results from their theoretical 
calculations.  N.R.K.\ would like to thank M.I.L.\ for many 
useful discussions.

This work was supported in part by grants from the Natural Science 
and Engineering Research Council of Canada and the National Science
Foundation (USA).

%




\end{document}